\documentclass[preprint,amsmath,showpacs]{revtex4}
\usepackage{graphicx}
\usepackage{dcolumn}
\usepackage{bm}

\def\bee{\begin{eqnarray}}
\def\eee{\end{eqnarray}}

\newcommand{\Deg}{{}^{\,\mbox{\scriptsize o}}}          

\begin{document}
\draft
\title{The energy spectrum of tau leptons induced by the high energy
Earth-skimming neutrinos}
\author{Jie-Jun Tseng$^{1,}$\footnote{E-mail: gen@phys.sinica.edu.tw},
 Tsung-Wen Yeh$^{2,}$\footnote{E-mail: twyeh@cc.nctu.edu.tw}, H. Athar$^{2,3,}$
 \footnote{E-mail: athar@phys.cts.nthu.edu.tw}, M. A.
 Huang$^{4,}$\footnote{E-mail: huangmh@phys.ntu.edu.tw},
Fei-Fain Lee$^{2,}$\footnote{E-mail: u8727515@cc.nctu.edu.tw},
 and Guey-Lin Lin$^{2,}$\footnote{E-mail: glin@cc.nctu.edu.tw}}
\address{$^{1}$ Institute of Physics, Academia Sinica,
Taipei 115, Taiwan\\
$^{2}$ Institute of Physics, National
Chiao-Tung University,
Hsinchu 300, Taiwan\\
$^{3}$ Physics Division, National Center for Theoretical Sciences,
Hsinchu 300, Taiwan\\
$^{4}$ Department of Physics, National Taiwan University, Taipei
10617, Taiwan}
\date{\today}
\begin{abstract}
We present a semi-analytic calculation of the tau-lepton flux
emerging from the Earth, induced by the incident high energy
neutrinos interacting inside the Earth for $10^{5} \leq
E_{\nu}/{\rm GeV} \leq 10^{10}$. We obtain results for the energy
dependence of the tau-lepton flux coming from the Earth-skimming
neutrinos, because  of the neutrino-nucleon charged-current
scattering as well as the resonant $\bar{\nu}_e e^-$ scattering.
We illustrate our results for several anticipated high energy
astrophysical neutrino sources such as the AGNs, the GRBs, and the
GZK neutrino fluxes. The tau lepton fluxes resulting from
rock-skimming and ocean-skimming neutrinos are compared. Such
comparisons can render useful information for the spectral indices of
incident neutrino fluxes.
\end{abstract}
\maketitle
\section{Introduction}
The detection of high energy neutrinos ($E_{\nu}> 10^{5}$ GeV)
 is crucial to identify
the extreme energy sources in the Universe, and possibly to unveil
the puzzle of cosmic rays with energy above the GZK cutoff\cite{Barwick:2000qp}.
 These proposed scientific aims are well
 beyond the scope of the conventional high energy gamma-ray
 astronomy. Because
 of the expected small flux of the high energy neutrinos, large
scale detectors ($\geq 1$ km$^{2}$) seem to be needed to obtain the
 first evidence.

 There are two different strategies to
detect the footprints of high energy neutrinos. The first
strategy is implemented by installing detectors in a large volume
of ice or water where most of the scatterings between the
candidate neutrinos and nucleons occur essentially inside the detector,
 whereas the second
strategy aims at detecting the air showers caused by the charged
leptons produced by the neutrino-nucleon scatterings taking place
inside the Earth or in the air, far away from the instrumented volume
 of the detector. The latter strategy thus include the possibility
of detection of quasi horizontal incident neutrinos which are also referred to
as the Earth-skimming neutrinos. These
  neutrinos are considered to interact below the horizon of an Earth based
surface detector.

The second strategy is proposed only
recently\cite{Domokos:1997ve}. The Pierre Auger observatory group
has simulated the anticipated detection of the
 air-showers from the decays of $\tau$
 leptons\cite{Bertou:2001vm}. The tau
air-shower event rates resulting from the Earth-skimming tau
neutrinos for different high energy neutrino
telescopes are given in\cite{Feng:2001ue}.
 A Monte-Carlo study of tau air-shower event rate was also reported not long
ago\cite{Bottai:2002nn}.
 We note that Ref.\cite{Feng:2001ue} does not consider the
tau-lepton energy distribution in the $\nu_{\tau}$-nucleon
scattering, and only the incident tau neutrinos with energies
greater than $10^{8}$ GeV are considered. For
Ref.\cite{Bottai:2002nn}, we note that only the sum of tau
air-shower event rates arising from different directions is given.
Hence some of the events may be due to tau-leptons/neutrinos
traversing a large distance.  As a result, it is not possible to
identify the source of tau-neutrino flux even with the observation
of tau-lepton induced air-shower.

In this work, we shall focus on the high energy Earth-skimming
neutrinos and shall calculate the energy spectrum of their induced
tau-leptons, taking into account the {\tt inelasticity} of
neutrino-nucleon scatterings and the tau-lepton {\tt energy loss}
in detail. Our work differs from
 Ref.\cite{Bottai:2002nn} by
our emphasis on the Earth-skimming neutrinos. We shall present our
results in the form of outgoing tau-lepton spectra for different
distances inside the rock, instead of integrating the energy
spectra. As will be demonstrated, such spectra are insensitive to
the distances traversed by the Earth-skimming $\nu_{\tau}$ and $\tau$.
They are essentially determined by the tau lepton range. Because
of this characteristic feature, our results are useful for
setting up simulations with specifically chosen air-shower content
detection strategy, such as detection of the Cherenkov radiation
or the air fluorescence.

We start with our semi-analytic description in Section II. The
transport equations governing the evolutions of neutrino and
tau-lepton fluxes will be derived. Using these, we then calculate
the tau-lepton flux resulting from the resonant  $\bar{\nu}_e
e^-\to W^-\to \bar{\nu}_{\tau}\tau^-$ scattering. In Section III,
we summarize our main results, namely the tau-lepton energy
spectra due to neutrino-nucleon scatterings. The implications of
our results will be discussed here also. In particular, we shall
point out that the ratio of tau-lepton flux induced by
rock-skimming neutrinos to that induced by ocean-skimming
neutrinos is sensitive to the spectral index of the incident
tau-neutrino flux. In Section IV, we discuss some prospects for
possible future observations of the associated radiation from
these tau leptons.

\section{Tau lepton energy spectrum}
Let us begin with the transport equations for tau neutrinos and
tau leptons. Considering only the neutrino-nucleon scatterings,
 we have
\bee
 \frac{\partial F_{\nu_{\tau}}(E,X)}{\partial
 X}=-\frac{F_{\nu_{\tau}}(E,X)}{\lambda_{\nu_{\tau}}(E)}+
 n_N\sum_{i=1}^{3}\int_{y^i_{\rm min}}^{y^i_{\rm max}}\frac{{\rm d}y}{1-y}
 F_i(E_y,X)\frac{{\rm d}\sigma^i_{\nu}}{{\rm d}y}(y,E_y),
 \label{transport1}
\eee
and
\bee
 \frac{\partial F_{\tau}(E,X)}{\partial
 X}&=&-\frac{F_{\tau}(E,X)}{\lambda^{\rm cc
 }_{\tau}(E)}-\frac{F_{\tau}(E,X)} {\rho d_{\tau}(E)}+
 \frac{\partial
 \left[(\alpha(E)+\beta(E)E)F_{\tau}(E,X)\right]}{\partial
 E}\nonumber \\
 &+& n_N\int_{y_{\rm min}}^{y_{\rm max}}\frac{{\rm d}y}{1-y}
 F_{\nu_{\tau}}(E_y,X)\frac{{\rm d}\sigma_{\nu_{\tau}N\to \tau Y
 }}{{\rm d}y}(y,E_y),
 \label{transport2}
\eee
where $n_N$ the number of target nucleons per unit medium mass,
and $\rho$ is the mass density of the medium. The
 $\sigma_{\nu}^{1,2,3}$ are defined as $\sigma(\nu_{\tau}+ N\to
 \nu_{\tau}+Y)$, $\Gamma(\tau \to \nu_{\tau}+Y)/c\rho n_N$, and
 $\sigma(\tau +N\to \nu_{\tau}+Y)$ respectively. The quantity $X$
represents the slant depth traversed by the particles, i.e., the
amount of medium per unit area traversed by the particle
 (and thus in units of g/cm$^{2}$). The
$\lambda_{\nu}$, $d_{\tau}$, and $\lambda^{\rm CC}_{\tau}$
represent the $\nu_{\tau}$ interaction thickness, the tau-lepton
decay length, and the tau-lepton charged-current interaction
thickness respectively, with, say, $\lambda_{\nu}^{-1}=n_N
\sigma_{\nu N }$ and $d_{\tau}=c\tau_{\tau}E/m_{\tau}$. The $E_y$
is equal to $E/(1-y)$,
 where $y$ is the inelasticity of neutrino-nucleon scatterings,  such that the
initial and final-state particle energies in the differential
cross sections ${\rm d}\sigma^i_{\nu}(y,E_y)/{\rm d}y$ and ${\rm
d}\sigma_{\nu_{\tau}N\to \tau Y }(y,E_y)/{\rm d}y$ are $E/(1-y)$
and $E$ respectively. The limits for $y$, $y_{\rm min}^i$ and
$y_{\rm max}^i$ depend on the kinematics of each process. Finally,
the energy-loss coefficients $\alpha (E)$ and $\beta(E)$ are
defined by $-{\rm d}E/{\rm d}X=\alpha (E)+\beta(E)E$ with $E$ the
energy being the tau lepton.
 An equation similar to Eq. (\ref{transport2}) in the context of
atmospheric muons was found in Ref.\cite{Volkova:af}.

As mentioned before, Eqs.~(\ref{transport1}) and
(\ref{transport2}) only take into account neutrino-nucleon
scatterings. It is of interest to calculate the tau-lepton
fluxes produced by the Glashow resonance\cite{shelly,Athar:2001bi}, namely
 via $\bar{\nu}_e e^-\to W\to \bar{\nu}_{\tau}\tau^-$ also. The
transport equation for $\bar{\nu}_e$ then reads:
\bee
\frac{\partial F_{\bar{\nu}_e}(E,X)}{\partial
 X}=-\frac{F_{\bar{\nu}_e}(E,X)}{\lambda_{\bar{\nu}_e}(E)}+
 n_N\int_{y_{\rm min}}^{y_{\rm max}}\frac{{\rm d}y}{1-y}
 F_{\bar{\nu}_e}(E_y,X)\frac{{\rm d}\sigma_{\bar{\nu}_e N\to \bar{\nu}_e
 Y}}{{\rm d}y}(y,E_y).
\label{transport3}
\eee
Similarly, the
corresponding equation for the tau-lepton flux is given by
\bee
 \frac{\partial F_{\tau}(E,X)}{\partial
 X}&=&-\frac{F_{\tau}(E,X)}{\lambda^{\rm cc
 }_{\tau}(E)}-\frac{F_{\tau}(E,X)} {\rho d_{\tau}(E)}\nonumber \\
 &+& n_e\int_{y_{\rm min}}^{y_{\rm max}}\frac{{\rm d}y}{1-y}
 F_{\bar{\nu}_e}(E_y,X)\frac{{\rm d}\sigma_{\bar{\nu}_ee^-\to
 \bar{\nu}_{\tau} \tau^- }}{{\rm d}y}(y,E_y),
\label{transport4} \eee where $n_e$ is the number of target
electrons per unit medium mass.

Before solving the above coupled transport equations, it is
essential to know the energy-loss coefficients $\alpha (E)$  and
$\beta(E)$. As pointed out before\cite{Dutta:2000hh}, the
coefficient $\alpha (E)$ is due to the energy loss by the
ionization\cite{Rossi}, while $\beta(E)$ is contributed by the
bremsstrahlung\cite{brem}, the $e^{+}e^{-}$ pair
production\cite{pair} and the photo-nuclear
processes\cite{Dutta:2000hh,Bezrukov:ci}. It is understood that
the contribution by $\alpha (E)$ becomes unimportant for
$E\geq 10^5$ GeV. The coefficient $\beta(E)$ can be
parameterized as $\beta(E)=(1.6+6(E/10^{9}{\rm GeV})^{0.2})\times
10^{-7} {\rm g}^{-1}{\rm cm}^2$ in the standard rock
 for $10^{5}\leq E/{\rm GeV} \leq 10^{12}$.

It is of interest to check the tau-lepton range given by our
semi-analytic approach. To do this, we rewrite
Eq.~(\ref{transport2}) by dropping the neutrino term, i.e.,
\bee
 \frac{\partial F_{\tau}(E,X)}{\partial
 X}=-\frac{F_{\tau}(E,X)}{\lambda^{\rm cc
 }_{\tau}(E)}-\frac{F_{\tau}(E,X)} {\rho d_{\tau}(E)}+
 \frac{\partial \left[\gamma(E)F_{\tau}(E,X)\right]}{\partial E},
\label{tau_alone} \eee
with $\gamma(E)\equiv \alpha (E)+\beta(E)E$. One can easily solve
it for $F_{\tau}(E,X)$:
\bee
 F_{\tau}(E,X)=F_{\tau}(\bar{E},0)\exp\left[\int_0^X
 {\rm d}T\left(\gamma^{\prime}(\bar{E})-
 \frac{1}{\rho d_{\tau}(\bar{E})}- \frac{1}{\lambda^{\rm
 cc}_{\tau}(\bar{E})}\right)\right],
\eee
where $\bar{E}\equiv
\bar{E}(X;E)$ with ${\rm d}\bar{E}/{\rm d}X=\gamma(\bar{E})$ and
$\bar{E}(0;E)=E$.
To calculate the tau-lepton range, we substitute
$F_{\tau}(E,0)=\delta(E-E_0)$. The survival probability $P(E_0,X)$
for a tau-lepton with an initial energy $E_0$ at $X=0$ is
\bee
 P(E_0,X)=\frac{\gamma(\tilde{E}_0)}{\gamma(E_0)}\exp\left[\int_0^X
 {\rm d}T\left(\gamma^{\prime}(\tilde{E}_0)-
 \frac{1}{\rho d_{\tau}(\tilde{E}_{0})}- \frac{1}{\lambda^{\rm
 cc}_{\tau}(\tilde{E}_0)}\right)\right],\label{prob}
\eee
where
$\tilde{E}_0\equiv \tilde{E}(X;E_0)$ with
${\rm d}\tilde{E}_0/{\rm d}X=-\gamma(\tilde{E}_0)$ and
$\tilde{E}_0(0;E_0)=E_0$. The tau-lepton range is simply
\bee
R_{\tau}(E_{0})=\int_{0}^{\infty}{\rm d}X P(E_0,X).
\label{range}
\eee
For $E_{0}=10^{9}$ GeV, we find that $R_{\tau}=10.8$ km in the
standard rock ($Z=11, \, A=22$) while $R_{\tau}=5.0$ km in the
iron. Both values are in good agreement with those obtained by
Monte-Carlo calculations\cite{Dutta:2000hh}. To compare the
tau-lepton ranges, we have followed the convention in
Ref.~\cite{Dutta:2000hh} by requiring the final tau-lepton energy
$\tilde{E}(X;E_0)$ to be greater than $50$ GeV.

\begin{figure}
\includegraphics[width=7.in]{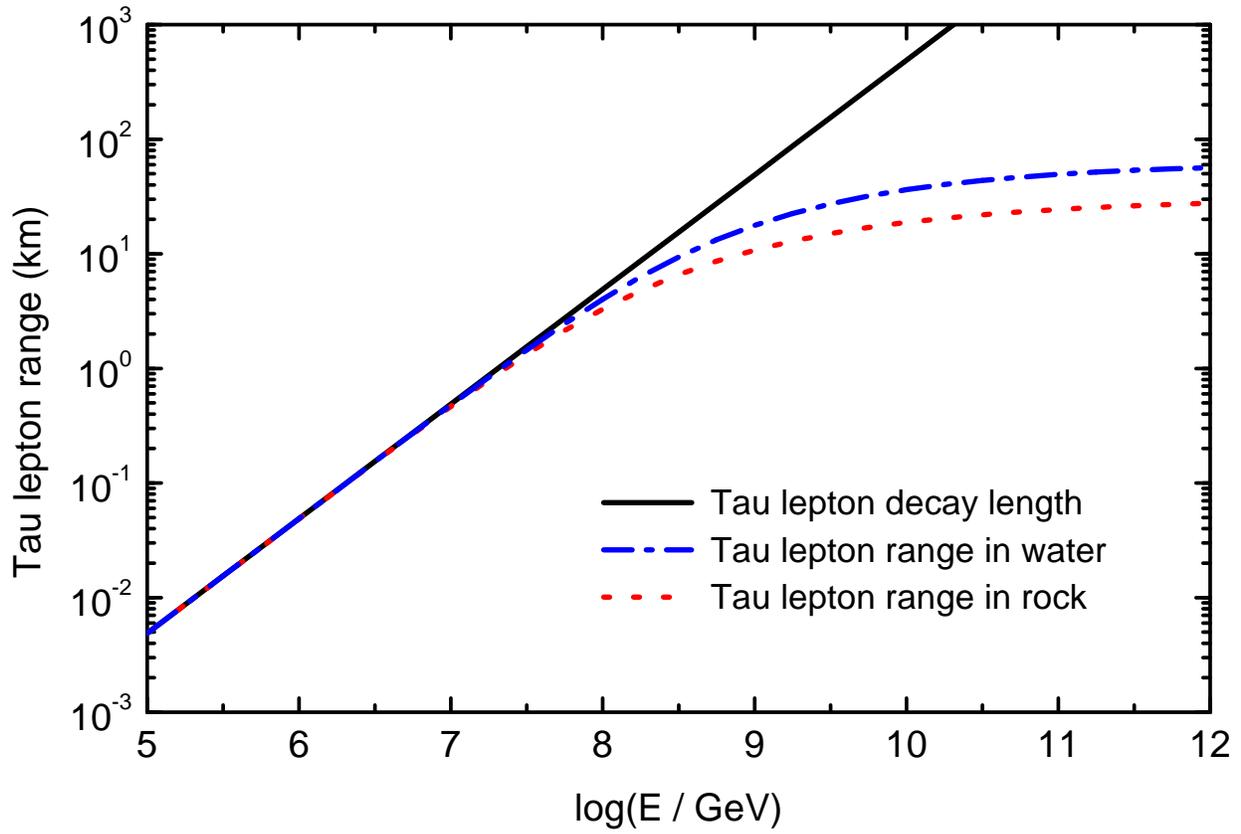}
\caption{The tau-lepton range in rock and in water using Eq.[\ref{range}]
 and the tau-lepton decay length $d_{\tau}$ in km as a function
 of tau-lepton energy in GeV.}
\label{figone}
\end{figure}

It is to be noted that we obtain $R_{\tau}$ by using the {\tt
continuous} tau-lepton energy-loss approach, rather than
stochastic approach adopted in Ref.\cite{Dutta:2000hh}. In the
muon case, the continuous approach to the muon energy loss is
known to overestimate the muon range\cite{Lipari:ut}. Such an
overestimate is not significant in the tau-lepton case, because of
the decay term in Eq.~(\ref{prob}). In fact, tau lepton decay term
dictates the tau range in the rock until $E_{\tau}\geq 10^7$ GeV.
Even for $E_{\tau}> 10^7$ GeV, the tau lepton range is still not
entirely determined by the tau-lepton energy loss. Hence different
treatments on the tau-lepton energy loss do not lead to large
differences in the tau-lepton range, in contrast to the case for
the muon range. Our result for the tau-lepton range up to
$10^{12}$ GeV are plotted in Fig. \ref{figone}. This is an
extension of the result in Ref.\cite{Dutta:2000hh}, where the
tau-lepton range is calculated only up to $10^{9}$ GeV. Our
extension is seen explicitly in the addition of charged-current
scattering term on the R.H.S. of Eq.~(\ref{tau_alone}). This term
is necessary because $1/\lambda_{\tau}^{\rm CC}$ becomes
comparable to $1/\rho d_{\tau}$ in the rock for $E\geq 10^{10}$
GeV; whereas one does not need to include the contribution by the
tau-lepton neutral-current scattering, since such a contribution
can not compete with the last term in Eq.~(\ref{tau_alone}) until
$E\geq 10^{16}$ GeV\cite{Dutta:2000hh}. We remark that our
extended results for $R_{\tau}$ are subject to the uncertainties
of neutrino-nucleon scattering cross section at high energies. We
use the CTEQ6 parton distribution functions\cite{Pumplin:2002vw}
in this work, and at the high energy (the small $x$ region, namely
for $x < 10^{-6}$), we  fit these parton distribution functions
into the form proportional to $x^{-1.3}$ as a guide.

Having checked the tau-lepton range, we now proceed to calculate
the tau-lepton flux. It is instructive to begin with the simple
case: the $\bar{\nu}_e e^-$ resonant scattering. It is well known
that\cite{shelly,Athar:2001bi}
\bee
 \sigma(\bar{\nu}_e e^-\to W^-\to \bar{\nu}_{\tau}\tau^-)=
 \frac{G_{F}^{2}m^{4}_{W}}{3\pi}\cdot \frac{s}{(s-m_W^2)^2+m_W^2\Gamma_W^2},
\eee
with $s=2m_{e}E_{\bar{\nu_{e}}}$ and $1/\sigma\cdot {\rm
d}\sigma/{\rm d}z=3(1-z)^{2}$,
 where $z= E_{\tau}/E_{\bar{\nu}_e}$. We shall only focus
on those $\bar{\nu}_e$'s for which $E_{\bar{\nu}_e}$ satisfies the
resonance condition, i.e., $E_{\bar{\nu}_e}\approx E_R\equiv
m_W^2/2m_e$.
 It is clear from Eq.~(\ref{transport4}) that
$F_{\tau}(E,X)$ only depends on $F_{\bar{\nu}_e}(E_R,X)$, because
of the narrow peak nature of $\bar{\nu}_e e^-$ scattering cross
section. One also expects that $F_{\tau}(E,X)$ is only significant
for $E$ around the resonance energy $E_R$. In this energy region,
one may neglect the first term on the R.H.S. of
Eq.~(\ref{transport4}) in comparison with the second term. In the
narrow width approximation, the last term in
Eq.~(\ref{transport4}) can be recast into
$\frac{1}{3}(1-E/E_R)^2(\pi \Gamma_W/L_R
m_W)F_{\bar{\nu}_e}(E_R,X)$, where $\Gamma_W$ is the width of the
$W$ boson while $L_R$ is the interaction thickness of the resonant
$\bar{\nu}_e e^{-}\to W^{-}$ scattering (see Appendix A for
details). The tau lepton flux can be readily obtained once
$F_{\bar{\nu}_e}(E_R,X)$ is given. We observe that the
regeneration term in Eq.~(\ref{transport3}) (second term on the
R.H.S.) can be neglected as it is necessarily off the $W$ boson
peak. Hence, we easily obtain
$F_{\bar{\nu}_e}(E_R,X)=\exp(-X/L_R)F_{\bar{\nu}_e}(E_R,0)$.
Substituting this expression into Eq.~(\ref{transport4}), we
obtain
\bee \frac{F_{\tau}(E,X)}{F_{\bar{\nu}_e}(E_R,0)}=3.3\cdot
 10^{-4}\times
 {\left(\frac{E}{E_R}\right)}\times \left(1-\frac{E}{E_R}\right)^2\times
 \exp\left(-\frac{X}{L_R}\right),\label{res_tau}
\eee
in the limit $X\gg \rho d_{\tau}$. The pre-factor $3.3\cdot
10^{-4}$ is obtained by assuming a standard-rock medium. In water
it becomes $1.4\cdot 10^{-4}$. It is to be noted that $E< E_{R}$
in the above equation. We shall see later that the contribution to
$F_{\tau}(E,X)$ by the $W$-resonance is negligible compared to
that by the $\nu_{\tau}-N$ scattering.

Let us now turn to the case of tau-lepton production by the
$\nu_{\tau}-N$ charged-current scattering. The tau-lepton flux can be
calculated from Eqs.~(\ref{transport1}) and (\ref{transport2})
once the incoming $\nu_{\tau}$ flux is given. The $\nu_{\tau}$
flux can be obtained by the following ansatz\cite{Naumov:1998sf}:
\bee
 F_{\nu_{\tau}}(E,X)=F_{\nu_{\tau}}(E,0)
 \exp\left(-\frac{X}{\Lambda_{\nu}(E,X)}\right),
\eee
where
$\Lambda_{\nu}(E,X)=\lambda_{\nu}(E)/\left(1-Z_{\nu}(E,X)\right)$,
with the factor $Z_{\nu}(E,X)$ arising from the regeneration
effect of the $\nu_{\tau}$ flux. On the other hand, the tau-lepton
flux is given by
\bee
 F_{\tau}(E,X)&=&\int_0^{X} {\rm d}T
 G_{\nu}(\bar{E},T)\nonumber \\
 &\times& \exp\left[\int_T^{X} {\rm d}T'\left(\gamma^{\prime}(\bar{E})-
 \frac{1}{\rho d_{\tau}(\bar{E})}- \frac{1}{\lambda^{\rm
 cc}_{\tau}(\bar{E})}\right)\right],\label{tau_flux}
\eee
with $\bar{E}\equiv \bar{E}(X-T;E)$, and
\bee
 G_{\nu}(E,X)=n_N\int_{y_{\rm min}}^{y_{\rm
 max}}\frac{{\rm d}y}{1-y}F_{\nu}(E_y,X)
 \frac{{\rm d}\sigma_{\nu_{\tau}N\to
 \tau Y }}{{\rm d}y}(y,E_y).
\label{source}
\eee
It is easy to see that the factor $Z_{\nu}(E,X)$  enters into
the expression for $F_{\tau}(E,X)$ through the function
$G_{\nu}(E,X)$. Similarly, $Z_{\nu}(E,X)$ also depends on
$F_{\tau}(E,X)$. It is possible to solve for $Z_{\nu}(E,X)$ and
$F_{\tau}(E,X)$ simultaneously by the iteration
method\cite{Naumov:1998sf}. The details are given in Appendix B.
\section{Results and discussion}
In the following, we show the tau-lepton fluxes resulting from
three kinds of diffuse astrophysical neutrino fluxes: the
AGN\cite{Kalashev:2002kx}, GRB\cite{Waxman:1997ti} and
GZK\cite{ber} neutrino fluxes. In these representative models,
$F_{\nu_{\tau}}$ arises because of neutrino flavor
mixing\cite{Athar:2002rr}. The $p\gamma $ interactions are the
source of intrinsic $F_{\nu_{\mu}}$, and $F_{\nu_{\tau}}=1/2\cdot
F_{\nu_{\mu}}$ because of (two) neutrino flavor oscillations
during propagation. Our convention for $F_{\nu_{\tau}}$ is that
$F_{\nu_{\tau}}={\mbox d}N_{\nu_{\tau}}/{\mbox d}(\log_{10}E)$ in
the unit of cm$^{-2}$ s$^{-1}$ sr$^{-1}$. The same convention is
used for the outgoing tau lepton fluxes.

\begin{figure}
\includegraphics{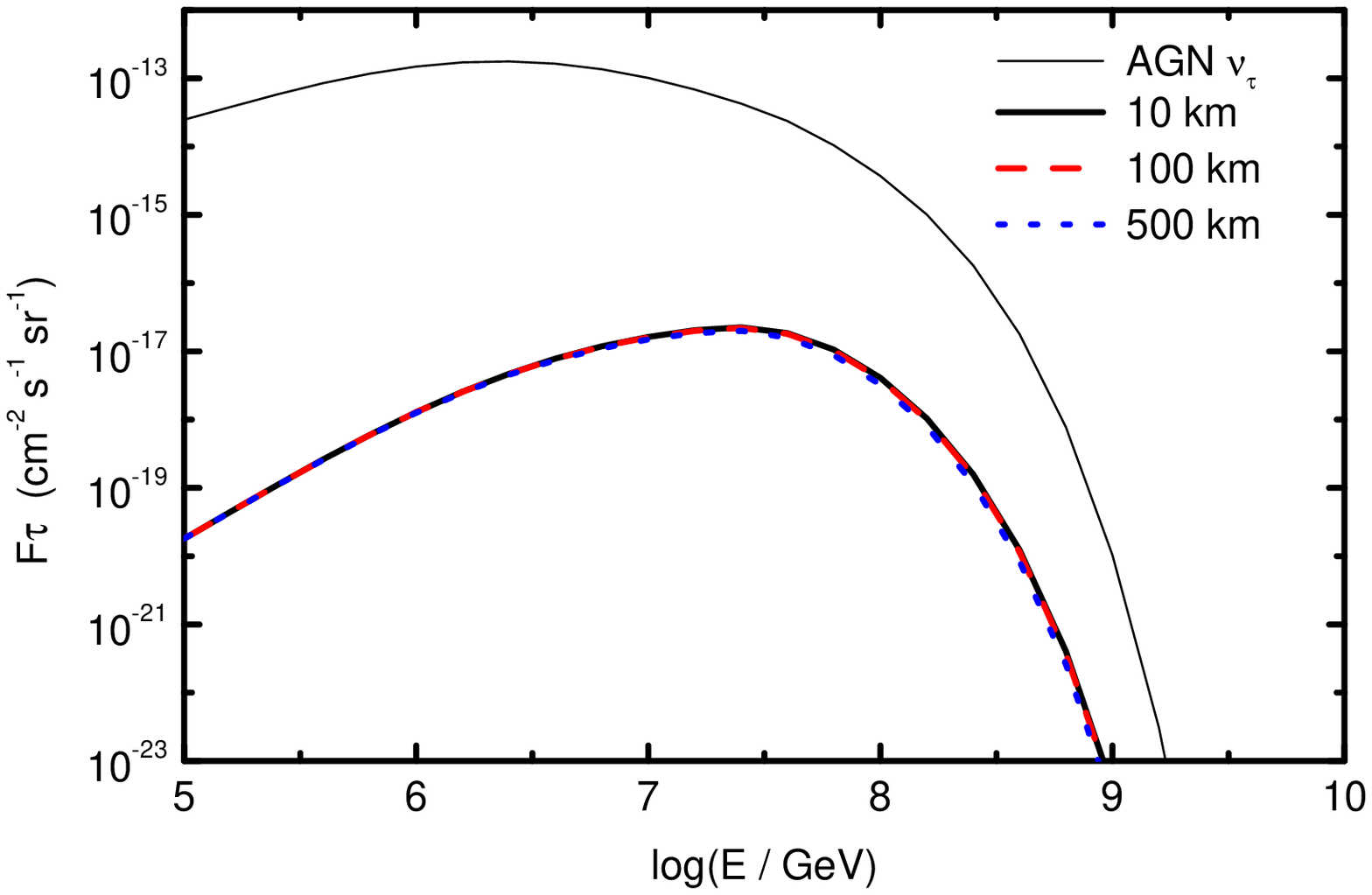}
\caption{The tau-lepton energy spectrum induced by the AGN neutrinos in rock  for three
 different $X/\rho$ ratio values (see text for more details). The incident tau-neutrino
 flux is shown by the thin solid line.}
\label{figtwo}
\end{figure}

\begin{figure}
\includegraphics{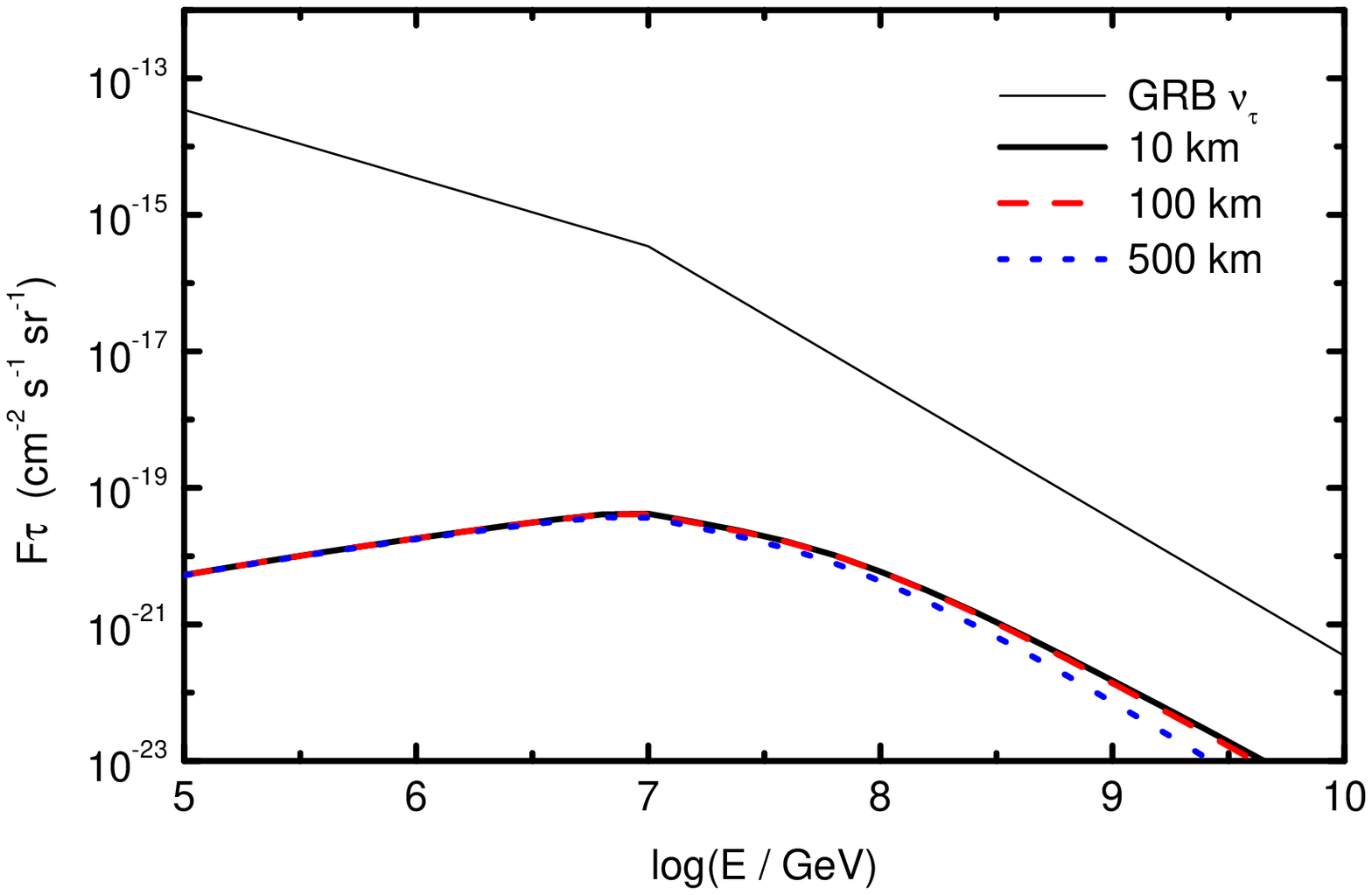}
\caption{The tau-lepton energy spectrum induced by the GRB neutrinos in rock  for three
 different $X/\rho$  ratio values (see text for more details). The incident tau-neutrino
 flux is shown by the thin solid line.}
\label{figthree}
\end{figure}

\begin{figure}
\includegraphics{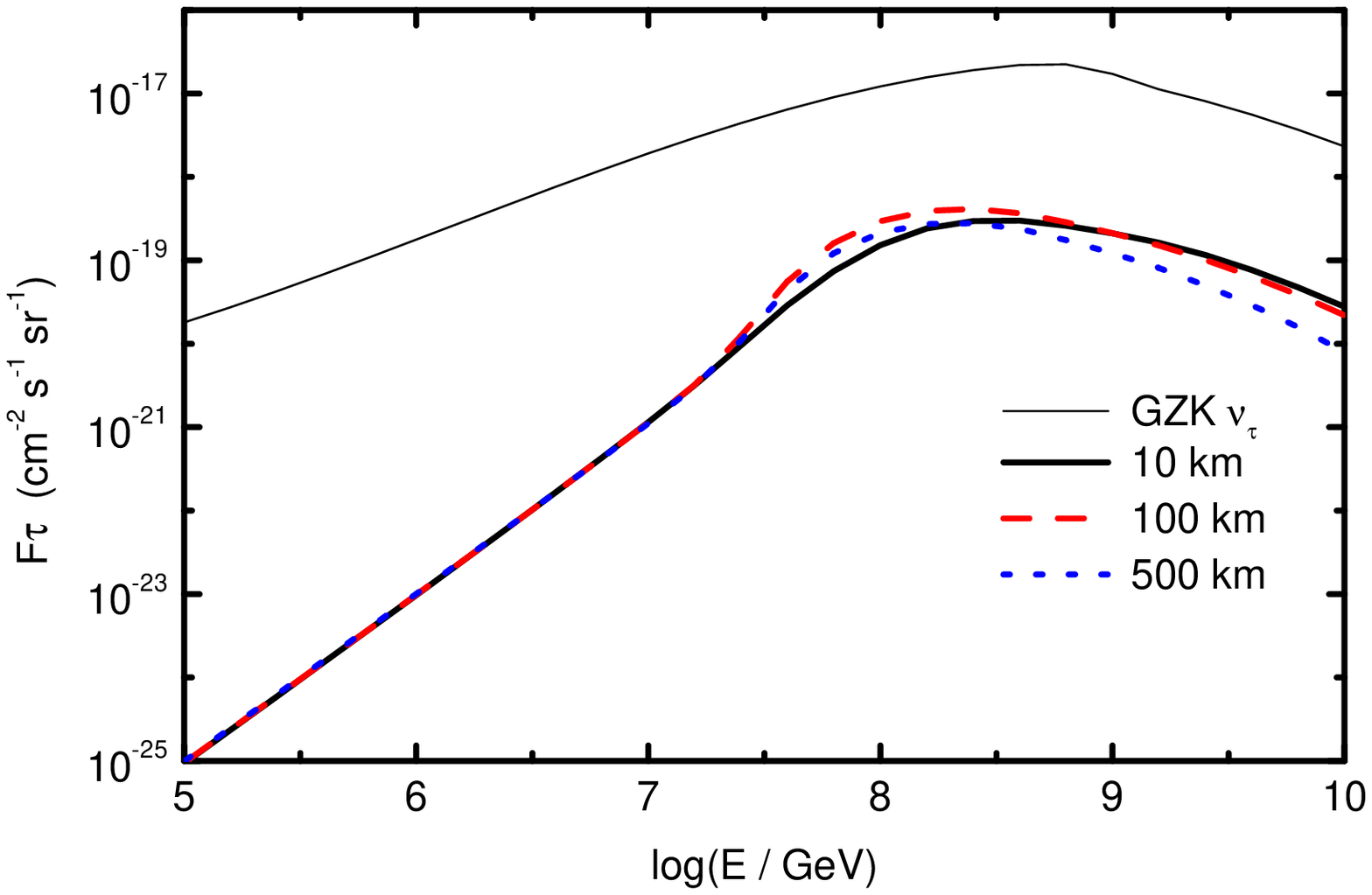}
\caption{The tau-lepton energy spectrum induced by the GZK neutrinos in rock for three
 different $X/\rho$ ratio values (see text for more details). The incident tau-neutrino
 flux is shown by the thin solid line.}
\label{figfour}
\end{figure}

In Fig. \ref{figtwo}, we show the outgoing tau-lepton energy
spectra resulting from the propagation of incident AGN neutrinos
inside the rock ($\rho=2.65 \ {\rm g/cm^{3}}$) for $X/\rho=10$ km,
$100$ km and $500$ km respectively. It is interesting to see that
the tau-lepton energy  spectra remain almost unchanged for the
above three different slant depths/matter density ratio values.
 This feature can be understood by two simple
facts. First of all, the neutrino-nucleon charged-current
interaction length, which is related to the interaction thickness
by $\lambda_{\rm CC}=\rho l_{\rm CC}$, is given by $l_{\rm
CC}=2\cdot 10^4 \ {\rm km} \ (\frac{1 \ {\rm
g/cm^{3}}}{\rho})(\frac{E_{\nu}}{10^{6}{\rm
 GeV}})^{-0.363}$. Secondly, the tau leptons, which eventually exit
the Earth, ought to be produced within a tau-lepton-range distance
to the exit point. For a tau-lepton produced far away from the
exit point, it loses energy and decays before reaching to the exit
point. Hence the tau-lepton flux is primarily determined by the
ratio of tau-lepton range to the charged current neutrino-nucleon
interaction length. The total slant depth $X$ which the
tau-neutrino (tau-lepton) traverses inside the Earth is then
unimportant, unless $X$ is large enough such that the tau neutrino
flux attenuates significantly before tau-neutrino is converted
into the tau-lepton. We note that the typical energy for the AGN
neutrinos, in which this flux peaks, is between $10^{5}$ and
$10^{8}$ GeV. The corresponding neutrino-nucleon neutral current
interaction length then ranges from $42,000$ km down to $3,400$
km, given $l_{\rm NC}=2.35 \cdot l_{\rm CC}$. Hence, even for
$X/\rho$ as large as $500$ km, the attenuation of the tau neutrino
flux is negligible. This explains the insensitivity of tau-lepton
flux with respect to our chosen $X/\rho$ values for the AGN case.
The situation is rather similar for the tau-lepton flux resulting
from the GRB tau neutrinos (see Fig. \ref{figthree}).
On the other hand, a slight suppression is found for the GZK case
at $E_{\tau}> 10^9$ GeV as one increases $X/\rho$ from $10$ km to
$500$ km (see Fig. \ref{figfour}).
 This is because  the typical GZK tau
neutrino flux peaks  for energy range between $10^{7}$ and
$10^{10}$ GeV, which corresponds to attenuation lengths ranging
from $7,800$ km down to $640$ km. One notices that $640$ km is
rather close to the distance $500$ km which we choose for
$X/\rho$. Hence a slight suppression on the tau-lepton flux occurs
for $X/\rho=500$ km.

We have compared our AGN-type tau-lepton flux with that obtained
by Monte-Carlo simulations, adopting stochastic approach for the
tau-lepton energy loss\cite{private}.  The two tau-lepton fluxes
agree within $\sim 10\%$. This is expected since the tau-lepton
range obtained by the above two approaches agree well, as pointed
out before. It is easily seen from Fig. \ref{figtwo} to Fig.
\ref{figfour} that the AGN case has a largest tau-lepton flux
between $10^{6}$ and $10^{8}$ GeV. Since the resonant
$\bar{\nu}_e-e^-$ scattering cross section peaks at
$E_{\nu}=6.3\cdot 10^{6}$ GeV, it is of interest to compare the
integrated tau-lepton flux resulting from this scattering to the
one arising from neutrino-nucleon scattering. For the former case,
we integrate the tau-lepton energy spectrum from $10^{6}$ GeV to
$6.3\cdot 10^{6}$ GeV, and obtain $\Phi^{\rm R}_{\nu}=0.08 \ {\rm
km}^{-2}{\rm sr}^{-1}{\rm yr}^{-1}$. For neutrino-nucleon
scattering, we find that $\Phi^{\rm CC}_{\nu}=2.2 \ {\rm
km}^{-2}{\rm sr}^{-1}{\rm yr}^{-1}$ by integrating the
corresponding tau-lepton energy spectrum from $10^{6}$ GeV to
$10^{7}$ GeV. The detailed results for $\Phi^{\rm CC}_{\nu}$ are
summarized in Table \ref{tableone}. The entries in the Table
entitled {\em full} are obtained using the $F_{\tau}$ obtained in
this work, whereas the approximated values entitled {\em approx}
are obtained by following the description given in
Ref.\cite{Feng:2001ue}, which uses a constant $\beta$ and a
constant inelasticity coefficient for $\nu_{\tau}N$ scattering. We
remark that the authors in Ref.\cite{Feng:2001ue} have taken $E$
to be greater than $10^{8}$ GeV. Hence the integrated fluxes in
the column {\em approx} with energies less than $10^8$ GeV are
taken as extrapolations. Thus, one should compare the two
integrated fluxes only for $E> 10^{8}$ GeV. One can see that the
two integrated fluxes seem to agree for $E> 10^8$ GeV.  Besides
the integrated fluxes for $E> 10^8$ GeV, we also obtain integrated
tau-lepton fluxes for $10^6\leq E/{\rm GeV}\leq 10^8$. It is
easily seen that, in this energy range, the integrated tau-lepton
flux from Earth-skimming AGN neutrinos is relatively significant .

\begin{figure}
\includegraphics{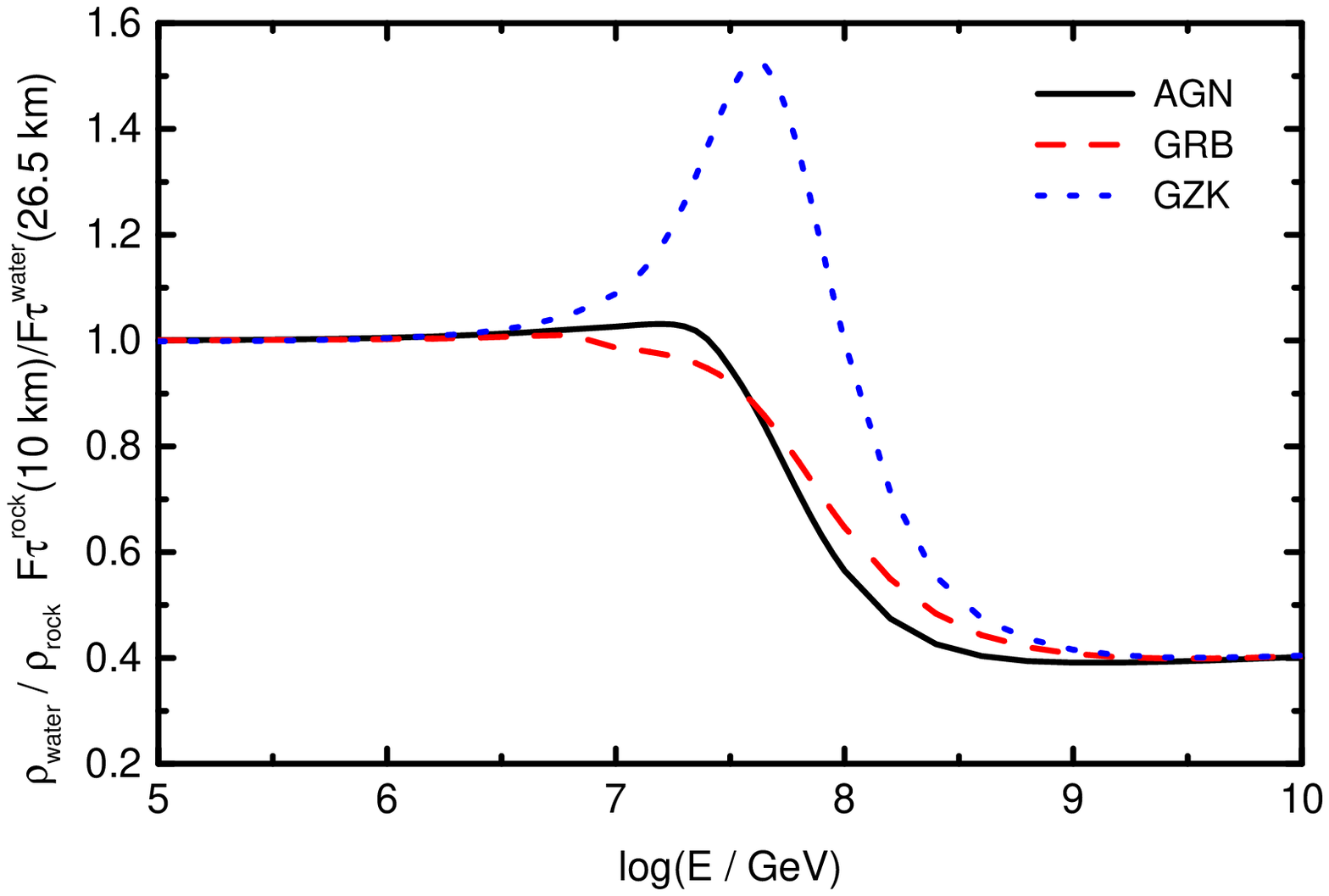}
\caption{The ratio of $F_{\tau}$ in rock and water induced by the AGN, the GRB and
 the GZK neutrinos for $X=2.65\cdot 10^{6}$ g/cm$^{2}$.}
\label{figfive}
\end{figure}

\begin{figure}
\includegraphics{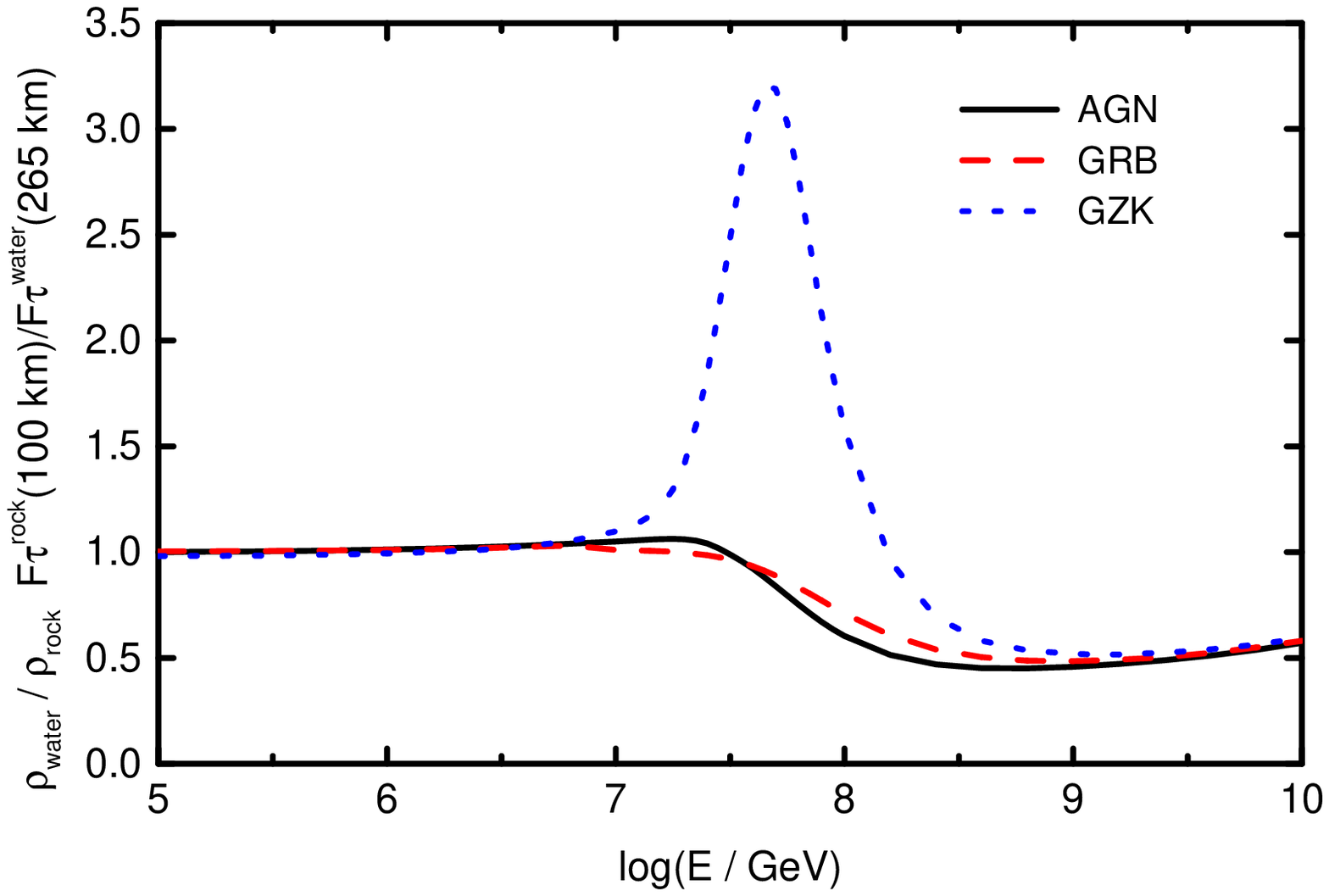}
\caption{The ratio of $F_{\tau}$ in rock and water induced by the AGN, the GRB and
 the GZK neutrinos for $X=2.65\cdot 10^{7}$ g/cm$^{2}$.}
\label{figsix}
\end{figure}

It is possible that the tau-neutrino skims through a part of the
ocean in addition to the Earth before exiting the interaction
region\cite{Hou:2002bh}. Hence, it is desirable to compare the
resulting tau-lepton fluxes as the tau neutrinos skim through
mediums with different densities, while the slant depths of
mediums are held fixed as an example.
 As stated before,  the tau-lepton flux is
essentially determined by the probability of $\nu_{\tau}N$
charged-current interaction happening within a tau-lepton range.
Furthermore, from Fig. \ref{figone}, it is clear that the
tau-lepton range equals to the tau-lepton decay length for
$E_{\tau}$ less than $10^{7}$ GeV. One therefore expects
$F_{\tau}^{\rm rock}(E,X)/F_{\tau}^{\rm water}(E,X)=\rho^{\rm
rock}/\rho^{\rm water}$ for $E_{\tau}< 10^{7}$ GeV. This is
clearly seen to be the case from Fig. \ref{figfive} and Fig.
\ref{figsix}, as we compare $F_{\tau}^{\rm rock}$ with
$F_{\tau}^{\rm water}(E,X)$ for $X=2.65\cdot 10^{6}$ g/cm$^{2}$
and $X=2.65\cdot 10^{7}$ g/cm$^{2}$ respectively. For $E_{\tau}>
10^{7}$ GeV, the tau-lepton range has additional dependencies on
the mass density and the atomic number of the medium. Hence the
ratio $F_{\tau}^{\rm rock}(E,X)/F_{\tau}^{\rm water}(E,X)$ starts
deviating from $\rho^{\rm rock}/\rho^{\rm water}$. It is
worthwhile to mention that the tau-lepton flux ratios for AGN and
GRB cases behave rather similarly. On the other hand, the ratio in
the GZK case has a clear peak between $10^{7.5}<E/{\rm
GeV}<10^{8.5}$. Such a peak is even more apparent for the slant
depth  $X=2.65\cdot 10^{7}$ g/cm$^{2}$. The appearance of this
peak has to do with the relatively
 flat
behavior of the incident GZK neutrino spectrum,
 while the position of this peak
is related to the energy dependencies of the tau-lepton range and
the neutrino-nucleon scattering cross sections.  We have confirmed
our observations by computing the flux ratios with simple
power-law incident tau-neutrino fluxes. The above peak in the
tau-lepton flux ratio implies the suppression of tau-lepton events
from ocean-skimming neutrinos compared to those from rock-skimming
neutrinos. As stated earlier, the suppression of ocean-skimming
neutrinos is related to the spectral index of the incident
neutrino flux. It is therefore useful to perform a detailed
simulation for it\cite{HLT}. Such a detailed study is needed
because the slant depths traversed by the above two kinds of
neutrinos are generally different.

\section{Prospects for possible future observations}

To observe the above tau leptons, the acceptance of a detector
must be of the order of $\sim \, {\rm km}^{2}{\rm sr}$. For AGN
neutrinos, the tau-lepton energy spectrum peaks at around $10^{7}$
to $10^{8}$ GeV, which is below the threshold of a fluorescence
detector, such as the
 High Resolution Fly's Eye (HiRes)\cite{Abu-Zayyad:2000ay}.
 Also, these tau leptons come
near horizontally. At present, it seems very difficult to construct a ground array in
vertical direction. A Cherenkov telescope seems to be a feasible solution.
 In this context, NuTel collaboration is developing Cherenkov
telescopes to detect the Earth-skimming high energy neutrinos\cite{Hou:2002bh}.
 However, because of the
small opening angle of Cherenkov light cone and only a 10\% duty
cycle (optical observations are limited to moonless and cloudless
nights only), such a detector must cover very large area and field
of view. A potential site for NuTel is at Hawaii Big Island, where
two large volcanos,
 namely Mauna Loa
and Mauna Kea, could be favorable candidates for high energy
 neutrinos to
interact with. For a detector situated on top of Mount Hualalai
and to look at both Mauna Kea and Mauna Loa, the required angular
field of view is $\sim 8\Deg \times 120\Deg$.
 Furthermore,
this telescope should have an acceptance area larger than 2 ${\rm
km}^{2}{\rm sr}$ so as to detect more than one event per year.

Concerning the GZK neutrinos, we note that the recent observation
of ultra high energy cosmic rays by HiRes seem to be consistent
with the GZK cutoff. Therefore a future observation of GZK tau
neutrinos shall provide a firm support to GZK cutoff. In
particular, the slight pile up of tau leptons between $10^{8}$ GeV
to $10^{9}$ GeV, induced by the Earth-skimming high energy GZK
neutrinos, should be a candidate signature for GZK neutrinos. The
integrated tau-lepton flux in this energy range is approximately
0.08 ${\rm km}^{-2}{\rm sr}^{-1}{\rm yr}^{-1}$. To detect one
event per year from this flux, the acceptance of a detector must
be larger than
 120 ${\rm km}^{2}{\rm sr}$, for a fluorescence detector (assuming a duty
cycle of 10\% ). Although HiRes can reach 1000 ${\rm km}^{2}{\rm
sr}$ at energy $> 3\cdot 10^9$ GeV, it would be a technical
challenge to lower down the threshold to $10^{8}$ GeV. Using a
system similar to HiRes, the Dual Imaging Cherenkov Experiment
(DICE)
  was able to detect Cherenkov light from
 extensive air-showers at energy as low as $10^{5}$ GeV\cite{Swordy:1999um}.
 However, the field of
view of DICE is also quite small, and thus several Cherenkov telescopes
would be needed.
 An alternative method is a hybrid detection of both Cherenkov and
fluorescence photons\cite{Cao}. That is, a detector similar to HiRes,
which looks at both
land and sea and detects both Cherenkov and fluorescence photons,
may observe the associated signal of GZK neutrinos.

In summary, we have given a semi-analytic treatment on the problem
of simultaneous propagation of high energy tau neutrinos and tau
leptons inside the Earth. Our treatment explicitly  takes into
account the {\tt inelasticity} of neutrino-nucleon scatterings as
well as the tau-lepton {\tt energy loss}. We specifically
considered the Earth-skimming situation and provided detailed
results for the energy dependencies of emerging tau-lepton fluxes
resulting from a few anticipated astrophysical neutrino fluxes.
The effect of matter density on the tau-lepton flux is also
studied. Such an effect is found to be related to the spectrum
index of incident neutrino flux. Our treatment thus provides a
basis for a more complete and realistic assessment of
high-energy-neutrino flux  measurements in the
under-construction/planning large neutrino telescopes.

\section*{Acknowledgements}
We thank N. La Barbera for communicating to us his
Monte-Carlo-based results. HA thanks Physics Division of NCTS for
support. MAH is supported by Taiwan's Ministry of Education under
{\it Research Excellence Project on Cosmology and Particle
Astrophysics: Sub-project II} with the grant number
92-N-FA01-1-4-2. FFL, GLL, JJT and TWY are supported by National
Science Council of Taiwan under the grant numbers
NSC91-2112-M009-019 and NSC91-2112-M-001-024.

\pagebreak
\appendix
\section{The contribution from resonant $\bar{\nu}_e e^-$ scattering}
The transport equations for $\bar{\nu}_e$ and the tau lepton are
given by Eqs.~(\ref{transport3}) and (\ref{transport4}). For
convenience, let us write $1-y=z$. The last term in
Eq.~(\ref{transport4}) can be simplified using
\begin{equation}
\frac{{\rm d}\sigma_{\bar{\nu}_ee^-\to
 \bar{\nu}_{\tau} \tau^- }}{{\rm d}z}(z,E/z)=\frac{m_W^4 G_F^2}{\pi}
 \frac{s(1-z)^2}{(s-m_W^2)^2+m_W^2\Gamma_W^2},
\end{equation}
and the narrow-width approximation
\begin{equation}
\frac{1}{\pi}\frac{m_W\Gamma_W}{(s-m_W^2)^2+m_W^2\Gamma_W^2}\approx
\delta(s-m_W^2).
\end{equation}
We arrive at
\begin{equation}
\frac{\partial F_{\tau}(E,X)}{\partial
 X}=-\frac{F_{\tau}(E,X)} {\rho d_{\tau}(E)}
 +\frac{1}{3}\left(1-\frac{E}{E_R}\right)^2
\left(\frac{\pi \Gamma_W}{L_R m_W}\right)F_{\bar{\nu}_e}(E_R,X),
\label{nuetau}
\end{equation}
where $E_R=m_W^2/2m_e$ is the $\bar{\nu}_e$ energy such that the
$W$ boson is produced on-shell in the $\bar{\nu}_e e^-$
scattering. The $L_R\equiv 1/n_e\sigma_{\bar{\nu}_e e^-\to W^-}$
is the interaction thickness for such a scattering. To solve for
$F_{\tau}(E,X)$, we need to input $F_{\bar{\nu}_e}(E_R,X)$.
Obviously, the $\bar{\nu}_e$ flux at the resonant-scattering
energy $E_R$ is mainly attenuated by the resonant scattering
itself. Hence
$F_{\bar{\nu}_e}(E_R,X)=\exp(-X/L_R)F_{\bar{\nu}_e}(E_R,0)$.
Substituting this result into Eq.~(\ref{nuetau}), we obtain
\begin{eqnarray}
F_{\tau}(E,X)&=&\frac{1}{3}\left(1-\frac{E}{E_R}\right)^2
\left(\frac{\pi \Gamma_W}{L_R
m_W}\right)F_{\bar{\nu}_e}(E_R,0)\exp\left(-\frac{X}{\rho
d_{\tau}(E)}\right) \nonumber \\
&\times&\int_0^X {\mbox d}Z\exp\left[\left(\frac{1}{\rho
d_{\tau}(E)}-\frac{1}{L_R}\right)Z\right].
\end{eqnarray}
The integration over $Z$ can be easily performed. In practice, it is
obvious that $X\gg \rho d_{\tau}(E)$. In this limit, we have
\begin{eqnarray}
F_{\tau}(E,X)=\frac{\pi}{3}\left(1-\frac{E}{E_R}\right)^2
\left(\frac{\Gamma_W}{m_W}\right) \left(\frac{ \rho
d_{\tau}(E)}{L_R}\right)F_{\bar{\nu}_e}(E_R,0)\exp\left(-\frac{X}{L_R}\right).
\end{eqnarray}
Let us consider standard rock as the medium for $\bar{\nu}_ee^-$
scattering, we then have $\rho/L_R=n_e\rho\sigma_{\bar{\nu}_e
e^-\to W^-}$. Given $\sigma_{\bar{\nu}_e e^-\to W^-}=4.8\cdot
10^{-31}$ cm$^2$ at the W boson mass peak, and $n_e\rho=2.65\times
6.0/2\times 10^{23}$/cm$^3$ in the standard rock, we obtain
$\rho/L_R=(26\, {\rm km})^{-1}$. Furthermore, we can write
$d_{\tau}(E)=49 \, {\rm km}\times (E/10^6{\rm GeV})$. We then
obtain the following ratio
\begin{eqnarray}
\frac{F_{\tau}(E,X)}{F_{\bar{\nu}_e}(E_R,0)}=3.3\cdot
 10^{-4}\times
 {\left(\frac{E}{E_R}\right)}\times \left(1-\frac{E}{E_R}\right)^2\times
 \exp\left(-\frac{X}{L_R}\right).
\end{eqnarray}
This is the result given by Eq.~(\ref{res_tau}) in the main text.
\pagebreak
\section{The iteration method for obtaining the $Z_{\nu}(E,X)$ and the $F_{\tau}(E,X)$}
The evolution for $F_{\nu_{\tau}}$ is given by
Eq.~(\ref{transport1}). With the ansatz
\bee
 F_{\nu_{\tau}}(E,X)=F_{\nu_{\tau}}(E,0)
 \exp\left(-\frac{X}{\Lambda_{\nu}(E,X)}\right),
\eee
we obtain the following equation for $Z_{\nu}(E,X)$:
\bee
XZ_{\nu}(E,X)&=&\int_0^X {\mbox d}X'\int_0^1 \frac{{\mbox
d}y}{1-y} \left\{
\frac{F_{\nu_{\tau}}^{(0)}(E_y)}{F_{\nu_{\tau}}^{(0)}(E)}\exp\left[-X'D_{\nu}(E,E_y,X')\right]
\Phi_{\nu_{\tau}}^{\rm NC}(y,E)
\right.\nonumber \\
&+&\frac{F_{\tau}(E_y,X')}{F_{\nu_{\tau}}^{(0)}(E)}
\left(\frac{\lambda_{\nu}(E)}{\rho d_{\tau}(E)}\right)
\exp\left(\frac{X'}{\Lambda_{\nu}(E,X')}\right)\Phi_{\tau}^d(y,E)\nonumber \\
&+&\left.\frac{F_{\tau}(E_y,X')}{F_{\nu_{\tau}}^{(0)}(E)}\left(\frac{\lambda_{\nu}(E)}
{\lambda_{\tau}(E)}\right)
\exp\left(\frac{X'}{\Lambda_{\nu}(E,X')}\right) \Phi_{\tau}^{\rm
CC}(y,E)\right\},\label{iterationZ}
\eee
 where
$F_{\nu_{\tau}}^{(0)}(E)\equiv F_{\nu_{\tau}}(E,0)$, while
$\Phi_{\nu_{\tau}}^{\rm NC}$, $\Phi_{\tau}^{\rm CC}$ and
$\Phi_{\tau}^d$ are respectively given by
\bee \Phi_{\nu_{\tau}}^{\rm NC}(y,E)=\frac{\sum_T
n_T\frac{\displaystyle {\mbox d}\sigma_{\nu_{\tau} T\to \nu_{\tau}
Y }}{\displaystyle {\mbox d}y}(y,E_y)}{\sum_T n_T
\sigma_{\nu_{\tau} T}^{\rm tot}(E)}, \eee
\bee
 \Phi_{\tau}^{\rm CC}(y,E)=\frac{\sum_T  n_T\frac{\displaystyle {\mbox
d}\sigma_{\tau T\to \nu_{\tau} Y }}{\displaystyle {\mbox
d}y}(y,E_y)}{\sum_T n_T \sigma_{\tau T}^{\rm tot}(E)}, \eee
\bee \Phi_{\tau}^d(y,E)=\frac{1}{\Gamma_{\tau}(E)}\frac{{\mbox
d}\Gamma_{\tau\to \nu_{\tau}Y}}{{\mbox d}y}(y,E_y), \eee
with $n_T$ the number of targets per unit mass of the medium, and
\bee
D_{\nu}(E,E_y,X)=\frac{1}{\Lambda_{\nu}(E_y,X)}-\frac{1}{\Lambda_{\nu}(E,X)}.
\eee
For the simplicity in notations, we take the lower and upper
limits for the $y$ integration to be $0$ and $1$ respectively. In
reality, the limits depend on the actual kinematics of each
process. One may impose these limits in the functions
$\Phi_{\nu_{\tau}}^{\rm NC}$, $\Phi_{\tau}^{\rm CC}$ and
$\Phi_{\tau}^d$.

To perform the iteration, we begin by setting $Z_{\nu(0)}=0$. In
this approximation, we have
\bee
F_{\nu_{\tau}(0)}(E,X)=F_{\nu_{\tau}}(E,0)
 \exp\left(-\frac{X}{\lambda_{\nu}(E,X)}\right).
\eee
Substituting $F_{\nu_{\tau}(0)}(E,X)$ into Eq.~(\ref{tau_flux}),
we obtain the lowest order $\nu_{\tau}$ flux, $F_{\tau(0)}(E,X)$.
The first iteration for $Z_{\nu}$, denoted by $Z_{\nu(1)}$ is
calculable from Eq.~(\ref{iterationZ}) by substituting
$F_{\nu_{\tau}(0)}(E,X)$, $F_{\tau(0)}(E,X)$, and $Z_{\nu(0)}$
into the R.H.S. of this equation. From $Z_{\nu(1)}$, we can then
calculate $F_{\nu_{\tau}(1)}(E,X)$ and $F_{\tau(1)}(E,X)$, which
corresponds to the results presented in this paper. We have checked
the convergence of iteration procedure and have found negligible
differences between $Z_{\nu(2)}$ and $Z_{\nu(1)}$ and their
associated $\nu_{\tau}$ and $\tau$ fluxes.

The value of $Z_{\nu}$ depends on the spectrum index of the
neutrino flux, since it effectively gives the regeneration effect
in the neutrino-nucleon scattering. In general, a flatter neutrino
spectrum implies a larger $Z_{\nu}$. The $Z_{\nu}$ is however not
sensitive to the slant depth $X$. In the case of GRB neutrinos,
where the flux decreases as $E_{\nu}^{-2}$ for $E_{\nu}< 10^7$
GeV, and decreases as $E_{\nu}^{-3}$ for energies greater than
that, we obtain $Z^{\rm GRB}_{\nu}\approx 0.2$. For the AGN
neutrino, $Z^{\rm AGN}_{\nu}$ changes from $0.96$ to $0.35$ as
$E_{\nu}$ runs from $10^5$ GeV to $10^6$ GeV. In this energy
range, the neutrino flux decreases slower than $E_{\nu}^{-0.5}$.
For $E_{\nu}$ greater than $10^8$ GeV, $Z^{\rm AGN}_{\nu}$ drops
below $0.2$ as the neutrino flux spectrum begins a steep fall. The
values for $Z^{\rm GZK}_{\nu}$ also follow the similar pattern.
\pagebreak
\begin{table}
\caption{Comparison of the integrated tau-lepton flux
 (km$^{-2}$yr$^{-1}$sr$^{-1}$) in different energy bins for the AGN, the GRB and the GZK neutrinos without
 and with approximation (see text for details).  The distance traversed is taken to be 10 km in rock here.
 For $10^{9}\leq  E/{\rm GeV} \leq   10^{10}$, the incident AGN neutrino flux is too small so that its induced tau-lepton flux is not shown.}
\begin{tabular}{|c|c|c|c|c|c|c|}
\hline
\hline
\label{tableone}
 {\em Energy Interval} & \multicolumn{2}{c|}{\em AGN} & \multicolumn{2}{c|}{\em GRB} &
 \multicolumn{2}{c|}{\em GZK} \\\cline{2-7}
 & full & approx &  full & approx &   full & approx   \\
 \cline{1-7}
  $10^{6}\leq  E/{\rm GeV} \leq   10^{7}$
 & 2.23 & 2.12 & $9.63\cdot 10^{-3}$  & $1.05\cdot 10^{-2}$ & $7.38\cdot 10^{-5}$  & $2.08\cdot 10^{-5}$\\
 $10^{7}\leq  E/{\rm GeV} \leq   10^{8}$
 & 4.89 & 5.12 & $7.12\cdot 10^{-3}$  & $6.82\cdot 10^{-3}$ & $1.14\cdot 10^{-2}$  & $1.90\cdot 10^{-2}$\\
$10^{8}\leq  E/{\rm GeV} \leq   10^{9}$
 & $1.95\cdot 10^{-1}$ & $1.52\cdot 10^{-1}$ & $5.39\cdot 10^{-4}$  & $4.63\cdot 10^{-4}$ & $8.17\cdot 10^{-2}$  &
 $8.47\cdot 10^{-2}$ \\
$10^{9}\leq  E/{\rm GeV} \leq   10^{10}$
 &   &  & $1.13\cdot 10^{-5}$  & $1.24\cdot 10^{-5}$ &  $3.31\cdot 10^{-2}$ & $3.52\cdot 10^{-2}$\\

\hline
\hline
\end{tabular}
\end{table}
\vspace*{5in}
\pagebreak
\pagebreak
\pagebreak
\pagebreak
\end{document}